\documentclass[12pt]{article}

\usepackage{scicite}
\usepackage{times}

\usepackage{bm} 
\usepackage{graphicx}

\usepackage{booktabs}       
\usepackage{bm}
\usepackage{amsmath}
\usepackage{xcolor}

\newenvironment{sciabstract}{%
\begin{quote} \bf}
{\end{quote}}

\title{A Simple Algorithm for Scalable Monte Carlo Inference}

\author
{Alexander Borisenko,$^{1}$ Maksym Byshkin,$^{2\ast}$ Alessandro Lomi$^{2}$\\
\\
\normalsize{$^{1}$National Science Center "Kharkiv Institute of Physics and Technology", Kharkiv, 61108, Ukraine}\\
\normalsize{$^{2}$Institute of Computational Science, Universit\`{a} della Svizzera italiana, Lugano, 6900, Switzerland}\\
\\
\normalsize{$^\ast$To whom correspondence should be addressed; E-mail:  maksym.byshkin@usi.ch}
}

\date{}

\begin{document} 


\baselineskip24pt


\maketitle


\begin{sciabstract}
The methods of statistical physics are widely used for modelling complex networks. Building on the recently proposed Equilibrium Expectation approach, we derive a simple and efficient algorithm for maximum likelihood estimation (MLE) of parameters of exponential family distributions - a family of statistical models, that includes Ising model, Markov Random Field and Exponential Random Graph models. Computational experiments and analysis of empirical data demonstrate that the algorithm increases by orders of magnitude the size of network data amenable to Monte Carlo  based inference. We report results suggesting that the applicability of the algorithm may readily be extended to the analysis of large samples of dependent observations commonly found in biology, sociology, astrophysics, and ecology. 
\end{sciabstract}


\section*{Introduction}

Statistical mechanics aims to describe and model large variety of complex systems. One example of complex systems is a social network, that consists of nodes representing social agents (e.g. individuals, groups, or organizations), and a set of social ties recording the presence of a relation among these agents. The Gibbs-Boltzmann distribution is one of the most fundamental formulas in statistical mechanics.  It allows to compute average behaviour of complex systems on the basis of their microscopic constituents. One of the first steps of statistical modelling is the development of a model that adequately describes the observed data. The difficulty in describing complex systems lies partly in limited efficiency of contemporary computing machines and computational methods. Statistical modelling is problematic when data of interest contain a large number of variables linked by complex structural dependencies. As an illustration, consider the following probability distribution defined on a multidimensional vector $x$ (assume discrete without loss of generality): 
\begin{equation} \label{eq:1}
\pi (x|\bm\theta )=\exp \left(\bm{\theta}^T\bm{g}(x)\right)/Z(\bm{\theta}),
\end{equation}
where $\bm{g}(x)$ is a vector of sufficient statistics, $\bm{\theta}=(\theta_1,\theta_2,..,\theta_L)$ is a vector of parameters and $Z(\bm{\theta})=\sum_{x}\exp \left(\bm{\theta}^T\bm{g}(x)\right)$ . This class of distributions is known as an exponential family~\cite{Bishop2006, Lehmann2006}.
This distribution may be written in the following equivalent form:
\begin{equation} \label{eq:2}
\pi (x|\beta\bm\theta )=\exp \left(-\beta E(x, \bm{\theta})\right)/Z(\beta \bm{\theta}),
\end{equation}
where $\beta=1/T$ is the inverse temperature and $E(x, \bm{\theta})=-\bm{\theta}^T \bm{g}(x)$ is the energy. $Z(\bm{\theta})$ is also known as a partition function. Particular models may be selected by specifying appropriate sufficient statistics $\bm{g}(x)$. For instance, for spin systems with spin variables $s_{i}$, the Ising model with two parameters may be specified by two statistics: $g_1(x)=-\sum_{<i,j>}s_{i} s_{j}$ and $g_2(x)=-\sum_{i}s_{i}$. Model selection is not a simple matter and we refer interested readers to available literature~\cite{Burnham2003,Ghahramani2015}.

The probability to observe empirical data $x_{\rm{obs}}$ is called likelihood, and its logarithm is given by:
\begin{equation} \label{eq:3}
l(\bm{\theta}|x_{\rm{obs}})=\bm{\theta}^T\bm{g}(x_{\rm{obs}})- \log(Z(\bm{\theta})).
\end{equation}
Once the model is selected, the parameters of the model may be estimated by maximizing the loglikelihood function~(\ref{eq:3}): ${\bm{\hat\theta}}_{\rm{MLE}}=\mathrm{argmax}_{\bm{\theta}}\left(l\left(\bm{\theta} | x_{\rm{obs}}\right)\right)$. Its gradient is given by: 
$d l(\bm{\theta}|x_{\rm{obs}})/d\bm{\theta}=\bm{g}(x_{\rm{obs}})-E_{\bm{\theta}}(\bm{g}(x))$,  its second derivative is negatively defined~\cite{Lehmann2006}, and thus  ${\bm{\hat\theta}}_{\rm{MLE}}$ is a solution of the following equation:
\begin{equation} \label{eq:MOM}
\bm{g}(x_{\rm{obs}})=E_{\bm{\theta}}\bm{g}(x),
\end{equation}
where $E_{\bm{\theta}}\bm{g}(x)=\sum_{x}\pi (x|\bm\theta )\bm{g}(x)$. If dimension of $x$ is more than about 100 variables, then $Z(\bm{\theta})$ is intractable, i.e.,  the number of possible  states is so large that  $Z(\bm{\theta})$ cannot be computed accurately. Computational methods for these problems are under active investigation and development. The current state of research in this field is reviewed elsewhere~\cite{Park2018,Nguyen2017,hunter2012computational}. The problem of parameter estimation via equation~(\ref{eq:MOM}) is very general and appears, among others, in astrophysics~\cite{Sharma2017}, computational biology and neuroscince~\cite{Nguyen2017,Beerli2005}, network science~\cite{Snijders2002,Stadtfeld2018,Decelle2011} and machine learning~ \cite{Tieleman2008,Asuncion2010,He2004,iba2003}.
Equation~(\ref{eq:MOM}) formulates the main problem of our study. If instead of one observation $x_{\rm{obs}}$ we have independent and identically distributed (iid) training sample $\left \{ x_j \right \}_{j=1}^{M}$, then ${\bm{\hat\theta}}_{\rm{MLE}}$ is a solution of: 
\begin{equation} \label{eq:MOM2}
\frac{1}{M}\sum_{j=1}^{M}\bm{g}(x_j)=E_{\bm{\theta}}\bm{g}(x),
\end{equation}
where in the LHS we have expectation under empirical data distribution and in the RHS we have expectation under the model distribution.

Traditionally, Monte Carlo methods~\cite{Newman1999,Robert2013} are used for the statistical inference on intractable statistical models. In general, Monte Carlo based methods provide asymptotically exact results, but they are computationally expensive and do not scale well to big data~\cite{Blei2017}. Numerous approximate methods were developed to overcome these problems of scale, but in many cases reliable, asymptotically exact methods are desirable. Markov chain Monte Carlo methods approximate $\pi (x|\bm\theta )$ and are used to compute  $E_{\bm{\theta}}\bm{g}(x)$. These methods appeared with the development of the Metropolis algorithm in the late 1940s. One step of the Metropolis algorithm~\cite{Metropolis1953} consists of: (i) proposing a random trial move $x\rightarrow x'$, and (ii) acceptance of this move with  probability: 
\begin{equation} \label{eq:Metropolis}
\alpha(x\rightarrow x', \bm{\theta})=\min \left\{1,e^ {\left[-\beta\left(E(x', \bm{\theta})-E(x, \bm{\theta})\right)\right]}\right\}.
\end{equation}
In 1970 Hastings~\cite{Hastings1970} proposed a simple but useful generalization of the Metropolis algorithm for a non-symmetric distribution $q(x\rightarrow x')$ of proposals. In this case the acceptance probability takes the form   
\begin{equation} \label{eq:Hastings}
\alpha(x\rightarrow x', \bm{\theta})=\min \left\{1, e^{\left[-\beta\left(E(x', \bm{\theta})-E(x, \bm{\theta})\right) \right]}\frac{q(x'\rightarrow x)}{q(x\rightarrow x')}\right\},
\end{equation}
and the transition probability is $P(x\rightarrow x')=q(x\rightarrow x')\alpha(x\rightarrow x', \bm{\theta})$.

While sampling from $\pi \left(x|\bm\theta \right)$ is a direct problem, the use of actual observations to infer the parameters of a model is an inverse problem. There are two popular approaches for computing the MLE. The first approach adapts the stochastic approximation~\cite{Robbins1951,Snijders2002} method to find the solution of~(\ref{eq:MOM}). The second was suggested by Geyer for the maximization problem~\cite{Geyer1991,geyer1992constrained}. These methods require simulation of many equilibrium configurations from $\pi \left(x|\bm\theta \right)$ and computation of expectations $E_{\bm{\theta}}\bm{g}(x)$. In 1988 Laurent Younes suggested an interesting algorithm~\cite{Younes1988} to compute the MLE. In machine learning it is known as a persistent contrastive divergence algorithm to train Restricted Boltzmann Machines~\cite{Tieleman2008,Tramel2018}. This algorithm is interesting because it finds the solution of~(\ref{eq:MOM}) without computing expectations $E_{\bm{\theta}}\bm{g}(x)$. In order to compute $E_{\bm{\theta}}\bm{g}(x)$, Monte Carlo simulation should be performed until convergence, while this convergence may be very slow. In contrast, the algorithm of Younes does not require converged Monte Carlo simulations between parameter updates. In the terminology of statistical physics, $x_{t}$ is a non-equilibrium system, which does not follow any stationary distribution $\pi \left(x|\bm\theta \right)$. The algorithm is given by the following parameter updating step: 
\begin{equation} \label{eq:Younes}
\bm\theta_{t+1}={\bm\theta}_{t}+a_{t}\left[\bm g\left(x_{\rm{obs}}\right) - \bm g\left(x_{t+1}\right)\right], 
\end{equation}
where $\Delta {\bm\theta}_{t}=a_{t}\left[\bm g\left(x_{\rm{obs}}\right) - \bm g\left(x_{t+1}\right)\right]$ is the step size, $x_{0}=x_{\rm{obs}}$ and $x_{t+1}$ is obtained from $x_{t}$ according to the method of the Gibbs sampler, more specifically by one step of the Gibbs sampler, and $a_{t}$ is decreasing learning rate, which decreases with  time $t$. Almost sure convergence to MLE was proved for a particular learning rate, but it was reported that it is impossible in practice to use such a small learning rate for which the convergence is proved. Younes wrote that in practice the starting point and the step size must be selected carefully. Recently, other authors confirmed these findings~\cite{Park2018}. In particular, Ib{\'a}{\~n}ez~\cite{iba2003} reported that “there can be significant differences of CPU time between good and bad choices. Even a bad choice can prevent the algorithm from converging”.

\section*{\label{Method}Method and Algorithm}

The peculiar feature of the algorithm~(\ref{eq:Younes}) is a simultaneous sampling and parameter adjustment strategy, rather than the common parameter adjustments based on the values of $E_{\bm{\theta}}\bm{g}(x)$. In Eq.~(\ref{eq:Younes}) $x_{t+1}$ is not an equilibrium sample of~(\ref{eq:1}), but is a function of both $\bm\theta$ and $x_{t}$.  Statistical mechanics may be helpful when nonequilibrium systems are considered. 

To the best of our knowledge, there are not many methods to perform sampling from probability distribution, when the parameters of these distribution are not constant. However, such methods exist and the comparison is helpful. One popular computational method, that bears similarity with statistical mechanics is simulated annealing. Simulated annealing was initially proposed as heuristic for multidimensional optimization. The essential idea of the method is simultaneous sampling and temperature decreasing strategy. This idea was generalized further by Mitsutake and Okamoto \cite{Mitsutake2009}. They suggested that instead of one dynamic parameter $\beta$, one can have many dynamic parameters, one coupling parameter for one physical quantity of interest. With these coupling parameters, Eq.~(\ref{eq:2}) may be written as 
\begin{equation} \label{eq:22}
\pi (x|\bm{\lambda \theta} )=\exp \left[ (\bm{\bm{\lambda  \theta}})^T\bm{g}(x)\right]/Z(\bm{\lambda \theta}),
\end{equation}
where vectors $\bm\lambda$ and $\bm\theta$ are multiplied elementwise. Thus it is possible to perform sampling and modify $\bm\lambda$ value simultaneously. For instance, for the Ising model, decribed in Introduction, one coupling parameter may be decreased during sampling, while the other one is constant. Or we can increase one coupling parameter, and discrease the other one simultaneously during sampling. 

Having established the relation of the algorithm ~(\ref{eq:Younes}) with statistical mechanics, we can try to improve this algorithm.  The efficiency of simulated annealing, parallel tempering and similar methods  \cite{Mitsutake2009} depends strongly on the choice of the temperature schedule. Kirkpatrick et al \cite{Kirkpatrick1983} proposed the following annealing schedule: $T_t=c^tT_0$, where $c<1$  and $1-c$ is small. Thus $T_{t+1}/T_t=c$, or equavalently $\beta_t/\beta_{t+1}=c$,  $(\beta\theta)_t/(\beta\theta)_{t+1}=c$. Hence $(\beta\theta)_{t+1}-(\beta\theta)_t=(1-c)(\beta\theta)_{t+1}$. Comparing to the algorithm (\ref{eq:Younes}), one has the following expression for the step size: $\Delta{(\beta\theta)}_{t}= (\beta\theta)_{t+1}-(\beta\theta)_t$, and the temperature schedule proposed by Kirkpatrick suggests that $\Delta{(\beta\theta)}_{t}=(1-c)(\beta\theta)_{t+1}$, where $1-c$ is small. Thus, the relation with statistical mechanics suggests that the step size in the algorithm ~(\ref{eq:Younes}) should be proportional to the corresponding parameter value. The generalization of Mitsutake and Okamoto may be straightforwardly applied to extend this idea to many parameters. Based on these considerations, we propose the following algorithm

\paragraph{Algorithm 1. Equilibrium Expectation algorithm.}

$x_{0}=x_{\rm{obs}}$  is given, 
\begin{equation} \label{eq:Max}
{\bm\theta}_{t+1}={\bm\theta}_{t}+a\cdot \max \left(\left| {\bm\theta}_{t} \right|, c\right)\cdot \mathrm{sign} \left[\bm g\left(x_{0}\right) - \bm g\left(x_{t+1}\right)\right],
\end{equation}
where $x_{t+1}$  is obtained from  $x_{t}$ by performing $m$ steps of the Metropolis-Hastings algorithm, all the operations are elementwise, $a$ is a learning rate, $c$ is a small positive constant to allow parameter values to change their sign.  Good starting point for this algorithm may be obtained by applying the contrastive divergence (CD) ${\bm\theta}_{0}\approx\bm{\hat\theta}_{\rm{CD}}(x_{obs})$ (see Supplementary materials). Initialization of the MLE algorithms by contrastive divergence was suggested by Hinton~\cite{Carreira-Perpinan2005} and Krivitsky~\cite{Krivitsky2017}.
MLE may be computed by averaging over the resulting ${\bm\theta}_t$ sequences:
\begin{equation} \label{eq:Max2}
{\bm{\hat\theta}}_{MLE}=\lim_{t \rightarrow \infty}\frac{1}{t-t_B}\sum_{j=t_B+1}^{t}{\bm\theta}_j,
\end{equation}
where $t_B$ is a burn-in time. The meaning of this averaging and the conditions when this algorithm computes MLE are given in Supplementary materials. A realization of this  algorithm in the pseudocode form is given in Supplementary materials.

If ${\theta}_{i}$ are not close to zero (formally, if $\left|\left \langle {\theta}_{i} \right \rangle \right|> c$ ) then the constant $c$ may be omitted and the parameter updating step~(\ref{eq:Max}) can be easily understood: if $g_{i}\left(x_{t+1}\right)< g_{i}\left(x_{0}\right)$  then we increase ${\theta}_{i}$  by 
$a\cdot \left|{\theta}_{i}\right|$,
and if $g_{i}\left(x_{t+1}\right)> g_{i}\left(x_{0}\right)$  then we decrease ${\theta}_{i}$  by 
$a\cdot \left|{\theta}_{i}\right|$.
Thus, up to a sign, the step size  $\Delta{\theta}_{i}$ is given by  $a\cdot{\theta}_{i}$.  Thus $\forall i,t : \Delta{\theta}_{i}/{\theta}_{i}=\pm{a} $ and if $a$ is small then for all the steps $t$ the step size is small \textit{relative} to the corresponding parameter value, and the model changes only slightly between parameter updates. 

The intuitive justification behind the algorithm given by Eq.~(\ref{eq:Younes}), is that it works because the parameter updates are small enough and thus the model changes only slightly between these updates ~\cite{Tieleman2008}. We can rely on the same intuition to explain the algorithm Eq.~(\ref{eq:Max}), but we argue that the model changes slightly between parameter updates if these parameter updates are small with respect to the current parameter values.

The EE algorithm can be used with different proposals $q\left(x\rightarrow x'\right)$.  Researchers in physics and statistics propose different samplers for different models, and the efficiency of these samplers depends on many different factors. Depending on the proposals  $q\left(x\rightarrow x'\right)$, a large number of samplers was developed, and the Gibbs sampler is one of them~\cite{Newman1999,Robert2013}.

\begin{figure*}[htb!]
	\centering
	\includegraphics[width=13.5cm,height=7.0cm]{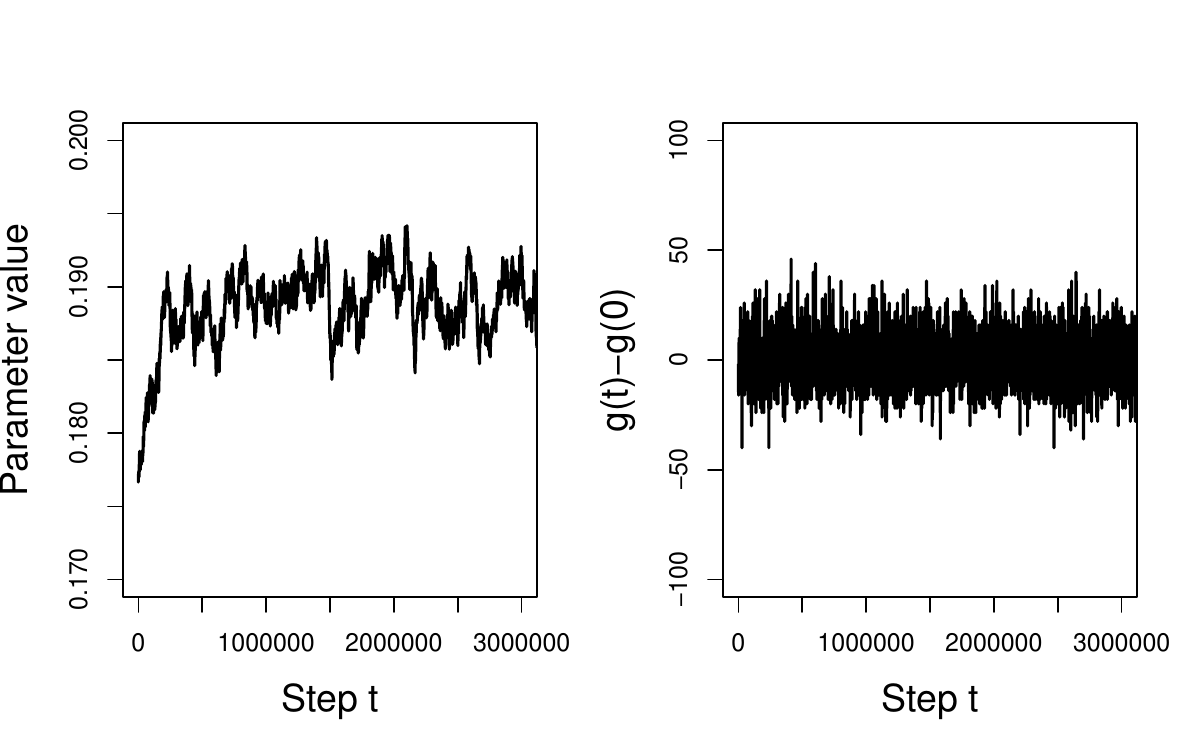}
	\caption{\label{Fig:2} Estimation of the Ising model parameter by the EE algorithm.}
\end{figure*}

\section*{Experimental analysis}

\subsection*{\label{Ising}A simple example}

To provide some intuition, we demonstrate the EE algorithm on the simplest Ising model with only one statistics $g_1(x)=-\sum_{<i,j>}s_i s_j$ and the corresponding interaction parameter  $\theta$. In all our computational experiments, unless stated otherwise, we used $m=1$, $c=0.01$ and the proposal $q\left(x\rightarrow x'\right)$ was defined as follows: one of the binary variables was selected randomly and uniformly and its value was changed to the opposite value. Here the learning rate was $a=0.001$, the empirical data $x_{\rm{obs}}$ was a small binary image with $8\times8$ pixels, see https://github.com/Byshkin/EquilibriumExpectation for details. The EE algorithm generated ${\theta}_{t}$ and $g_{1}\left(x_{t}\right) -  g_{1}\left(x_{\rm{obs}}\right)$ sequences, that we present in Fig.~\ref{Fig:2}. Here ${\theta}_{0}$  is a CD estimator for $x_{\rm{obs}}$.  ${\theta}_{0}=0$ is also possible, but increases the convergence time by several times, and we used CD as a starting point an all our experiments. Fig.~\ref{Fig:2} shows that $g_{1}\left(x_{t}\right) -  g_{1}\left(x_{\rm{obs}}\right)$ fluctuates around zero, and  ${\theta}_{t}$ fluctuates and converges starting from the step $t_B\approx 10^6$.

Using Eq.~(\ref{eq:Max2}), we obtain $\hat\theta_{\rm{MLE}}=0.189$.
A robust test that the estimated parameters values is a solution of Eq.~(\ref{eq:MOM}) was suggested by Snijders ~\cite{Snijders2002} and we checked that this convergence test is satisfied. 
In practice, it is useful to consider the following convergence test: i) the average of ${\bm\theta}_{t}$ converges to a constant according to Eq.~(\ref{eq:Max2}) and  ii) the following t-ratio test is satisfied: 
\begin{equation} \label{eq:t-ratio}
\frac{\left| \left \langle g_{i}\left(x_{t}\right)- g_{i}\left(x_{\rm{obs}}\right)\right \rangle \right|}{\sigma\left[g_{i}\left(x_{t}\right)- g_{i}\left(x_{\rm{obs}}\right)\right]} <\tau~  \forall i,
\end{equation}
where  $\left \langle .. \right \rangle$ and $\sigma(..)$ are the mean and the standard deviation over  $t>t_B$ ,  $\tau=0.1$. 
A relation between this convegrence tests and that used by Snijders is discussed in Supplementary materials.

\subsection*{\label{VBM}Fully visible Boltzmann machines and inverse Ising problem}

Now we consider the Visible Boltzmann Machine (VBM) model in the form
\begin{equation} \label{eq:VBM}
\pi_{VBM} (x|\bm{\theta} )=\frac{1}{Z(\bm{\theta})}\exp \left(-\frac{1}{2}\sum_{i, j}\theta_{ij}x_{i}x_{j}\right),
\end{equation}
where $x$ is a vector of 15 binary variables $x_{i}=\pm 1$ , $\theta_{ij}$  is a symmetric matrix of the model parameters and the partition function is
\begin{equation} \label{eq:partition}
Z(\bm{\theta} )=\sum_{x}\exp\left(-\frac{1}{2}\sum_{i, j}\theta_{ij}x_{i}x_{j}\right),
\end{equation}
where the summation runs over all $2^{15}$  states of the vector $x$.

\begin{figure*}
	\centering
	\includegraphics[width=13.0cm,height=6.0cm]{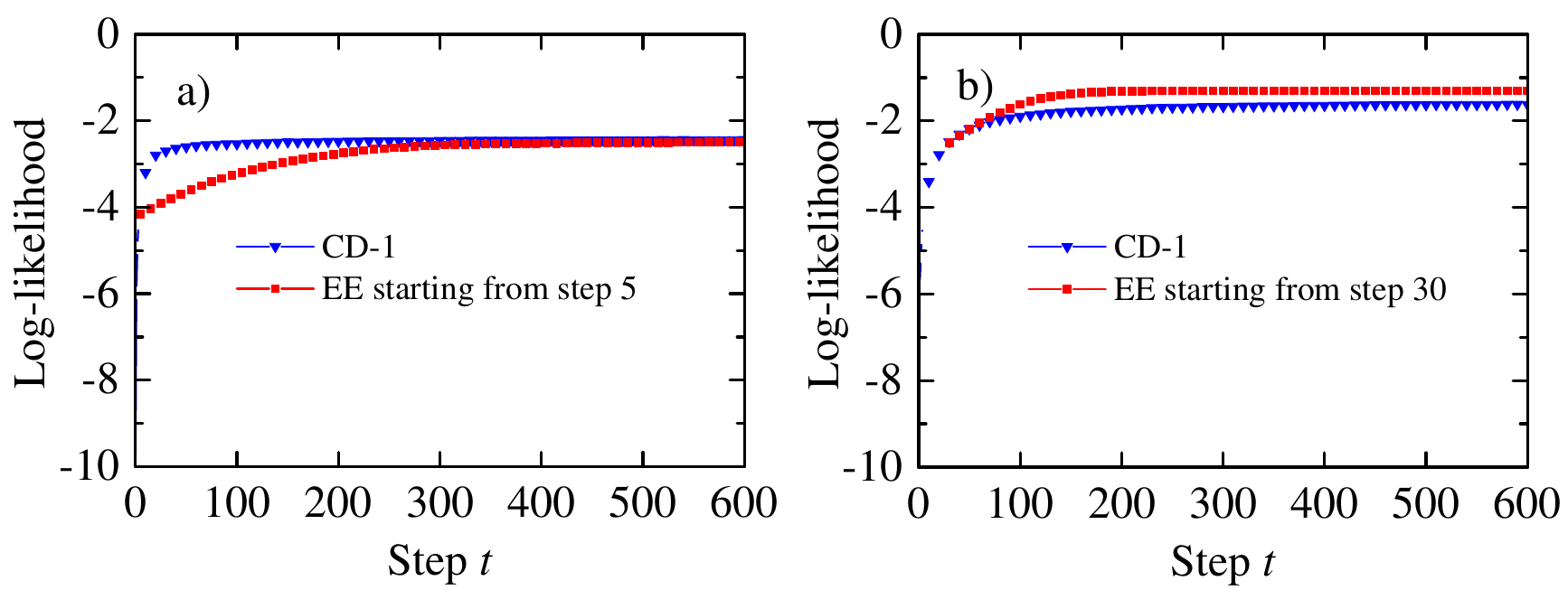}
	\caption{\label{Fig:2} \label{Fig:3} Log-likelihood~(\ref{eq:3}) calculated as a result of fitting of the observed data generated by the VBM model~(\ref{eq:VBM}) using a) VBM model~(\ref{eq:VBM}) and b) 1D Ising model~(\ref{eq:Ising}).}
\end{figure*}

We perform our experiment using an ensemble of $n=1000$ Markov chains in the following manner. At first we generate 105 parameters $\theta_{ij}\sim \mathcal{N}(0, 1)$ and anneal all $n$ chains for $10^{5}$  MC steps. Then we compute ensemble-averaged statistics
\begin{equation} \label{eq:g}
g_{ij}=\frac{1}{n}\sum_{k=1}^{n}x_{i}^{k}x_{j}^{k},
\end{equation}
which are used in our inference procedure. After that we set ${\theta}_{ij}=0$  and start the inference algorithms.

Our goal is to compare the performance of the CD and EE algorithms. For this purpose, we perform the inference procedure in two different ways. In the first experiment, we make inference of the entire $\theta_{ij}$ matrix, thus fitting the observed data~(\ref{eq:g}) with the VBM model. In the second experiment we fit the same data with the 1-D Ising model, keeping nonzero only nearest-neighbor matrix elements ${\theta}_{i, i\pm1}$  and applying periodic boundary conditions:
\begin{equation} \label{eq:Ising}
\pi_{1D-Ising} (x|\bm{\theta} )=\frac{1}{Z(\bm{\theta})}\exp \left(-\frac{1}{2}\sum_{\left|i-j\right|=1}\theta_{ij}x_{i}x_{j}\right).
\end{equation}

In both cases we start inference with the CD algorithm~\cite{Carreira-Perpinan2005} (learning rate $a=0.1$ and $m=1$) and after several initial steps start the EE algorithm (learning rate $a=0.005$ , $c=0.001$ and $m=1$). To visualize our results, we directly compute the log-likelihood function, given by Eq.~(\ref{eq:3}), and present its value as a function of the step $t$ in Fig.~\ref{Fig:3}.

Fig.~\ref{Fig:3}~(a) shows the calculated log-likelihood~(\ref{eq:3}) of the ensemble as a result of fitting the observed data using the VBM model~(\ref{eq:VBM}). The EE algorithm is initialized with ${\theta}_{ij}$ values calculated at step 4 of the CD algorithm. One can see that the convergence of the EE algorithm is slower than that of CD. After convergence, the results produced by CD and EE algorithms are equivalent. This computational result confirms our theoretical finding: CD is a consistent estimator\cite{Byshkin2018} (see Supplemetary Infromation). Furthermore, these results suggest that estimation of parameters by contrastive divergence is equivalent to estimation of parameters by MLE, when the data under study is fitted by the same model, by which it was genetated. A different situation is shown in Fig.~\ref{Fig:3}~(b). Here we show the likelihood of the data generated by the VBM model~(\ref{eq:VBM}) and fitted with the 1D Ising model~(\ref{eq:Ising}). Now the EE algorithm is initialized with ${\theta}_{ij}$ values calculated at step 29 of the CD algorithm. One can see that CD and EE algorithms converge to different values of the likelihood. The EE algorithm computes MLE and the likelihood obtained with this algorithm is significantly higher than that obtained with CD.

\subsection*{\label{MRF}Conditional random field}

\begin{figure*}
	\centering
	\includegraphics[width=13.5cm,height=5.5cm]{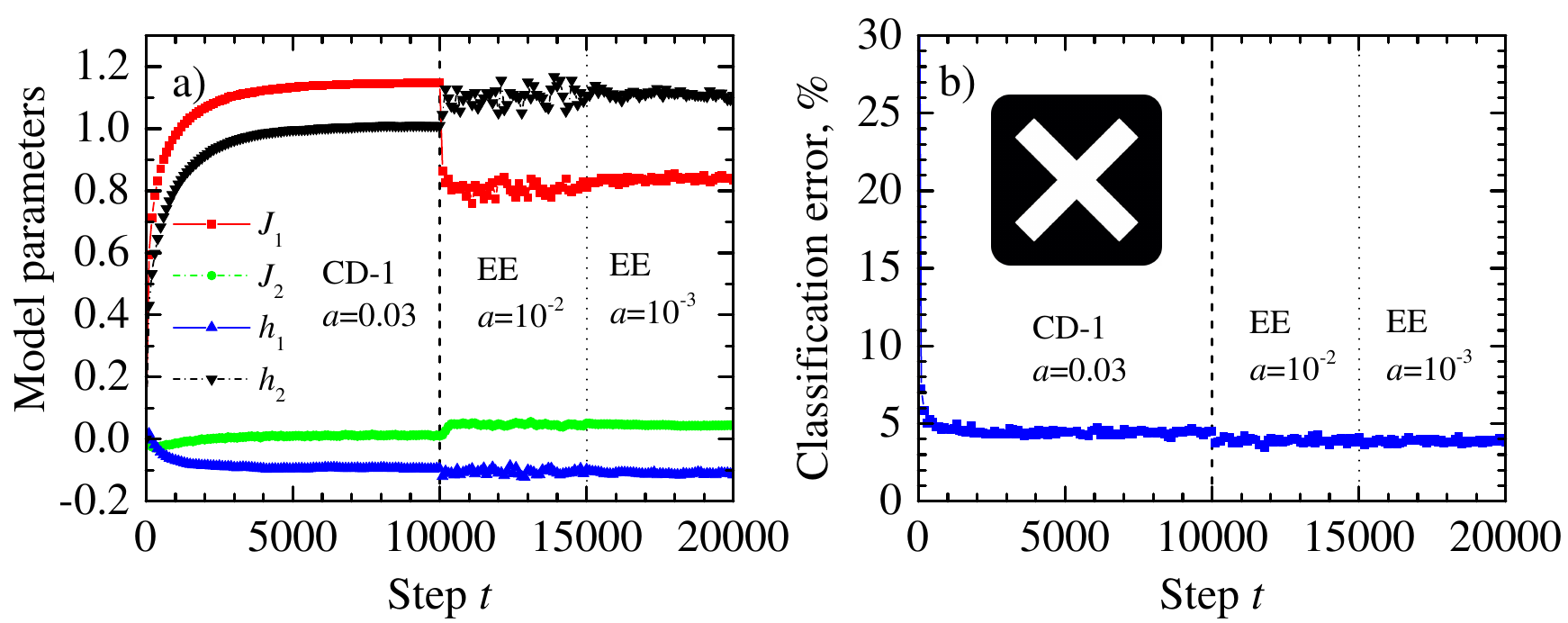}
	\caption{\label{Fig:4} CRF model parameters (a) and classification error~(\ref{eq:error}) (b) vs. number of steps $t$. The original  X-shaped image is shown  in the inset.}
\end{figure*}

We also test the EE algorithm on a conditional random field (CRF) model for image 
processing ~\cite{Asuncion2010}. Let $x$ be a binary image, where $x_{j}=\pm 1$  is the label of the $j$-th pixel. Let $y$ be a noisy observation of $x$. The conditional probability of pixel labels is given by:
%

\begin{equation}
\label{eq:CRF}
\pi_{CRF} (x|y, h, J)=
\frac{1}{Z(h, J)}e^{ \left[-\sum_{j} h^{T} f_{j}\left(y\right)x_{j}-\frac{1}{2}\sum_{i \sim j}J^{T} f_{ij}\left(y\right)x_{i}x_{j}\right]},
\end{equation}

where the notation $i \sim j$  indicates that the pixels $i$ and $j$ are nearest neighbors in the image, the vectors $f_{j}\left(y\right)=\left[1, y_{j}\right]$  and $f_{ij}\left(y\right)=\left[1, \left| y_{i}-y_{j}\right|\right]$  are called node features and edge features respectively and the vectors $h=\left[h_{1}, h_{2}\right]$  and $J=\left[J_{1}, J_{2}\right]$  are the model parameters. 
In our experiment we take a simple binary X-shaped image (see the inset in Fig.~\ref{Fig:4}~(b)) with dimensions $40\times40$ pixels and generate 10 learning samples and 5 testing samples by adding $\mathcal{N}(0, 1)$  noise to each pixel label.
%

We start the inference procedure by initializing CD with ${h}=\left[0, 0\right]$  and ${J}=\left[0, 0\right]$, and making $10000$ steps of the CD algorithm with the learning rate $a=0.03$. After that we run the EE algorithm with $m=1$ and $c=0.001$ for $5000$ steps with $a=0.01$  and next $5000$ steps with $a=0.001$. At each step of the CD and EE algorithms we use the obtained ${h}$ and ${J}$ values to anneal the test samples for 500 steps to calculate the classification error (the fraction of false pixels):
\begin{equation} \label{eq:error}
\mathrm{Error}=\frac{1}{2\cdot N_{\mathrm{test}} \cdot n_{\mathrm{pix}}}\sum_{k=1}^{N_{\mathrm{test}}}\sum_{i=1}^{n_{\mathrm{pix}}} \left| x_{i}^{k}-x_{i}^{\mathrm{orig}}\right|.
\end{equation}
Here $N_{\mathrm{test}}=5$, $n_{\mathrm{pix}}=1600$  is a total number of the image pixels and $x^{\mathrm{orig}}$  is the original image. In Fig.~\ref{Fig:4}~(a) and (b) we show the time dependence of the model parameters and the classification error~(\ref{eq:error}), respectively.

From Fig.~\ref{Fig:4}~(a) one can see that the CD estimates of parameters are significantly different from MLE, computed with the EE algorithm. This difference, however, has a small effect on the classification error (see Fig.~\ref{Fig:4}~(b)). The resulting value of the classification error agrees well with the results reported in Ref.~\cite{Asuncion2010}. Smaller classification error may be obtained using more advanced CRF specifications~\cite{He2004}.

\subsection*{\label{ERGM}Exponential random graph models}

Exponential random graph models (ERGMs) are widely used for the analysis of social~\cite{Lusher2013}, biological~\cite{Byshkin2018}, and brain networks~\cite{Simpson2011}. Very often these models are used for hypothesis testing: if an estimated parameter $\theta_i$ is significantly larger than zero then the corresponding $g_i(x_{obs})$ is larger than might be expected by chance, conditional on all the other parameters of the model. In case of ERGMs, $x$ is a vector of binary variables (0/1), denoting the  absence/presence of ties between network nodes. For directed networks the dimension of $x$ is $N \times (N-1)$, where $N$ is the number of network nodes. In contrast to other models that we consider in this paper, ERGMs do not belong to the class of Markov Random Field distributions and this fact complicates the problem of parameter estimation~\cite{Moores2015}. 

Recently Byshkin et al. proposed a fast adaptive algortihm to compute the MLE ~\cite{Byshkin2018}. Similar to the algorithm ~(\ref{eq:Younes}), the algorithm did not requiere converged Monte Carlo simulations between parameter updates. Though ERGMs are popular statistical models, there were no successful attempts of applying the algorithm ~(\ref{eq:Younes}) to ERGMs. The adaptive algorithm was successfully applied to compute MLE of ERGM parameters, and significantly increased the size of networks for which MLE may be computed ~\cite{Byshkin2018}. However, it required adaptation of learning rates for each parameter separately, and a complicated adaptive method was applied. Though  good empirical results were observed, the authors could not understand or explain their algorithm.

The algorithm ~(\ref{eq:Max}) does not requiere any adaptation. Good scaling properties of the EE algorithm can be easily understood: only one step of the Metropolis-Hastings algorithm is enough for one step of the EE algorithm.  For comparison,  one step of the stochastic approximation~\cite{Robbins1951,Snijders2002} requires a number of the Metropolis-Hasting steps which is larger than the burn-in time, and the burn-in time grows with $N$. Good scaling properties of the EE algorithm allow to fit ERGMs to complex networks with hundreds of thousands of nodes and billions of tie variables correspondingly.

\begin{figure*}
	\centering
	\includegraphics[width=13.5cm,height=7cm]{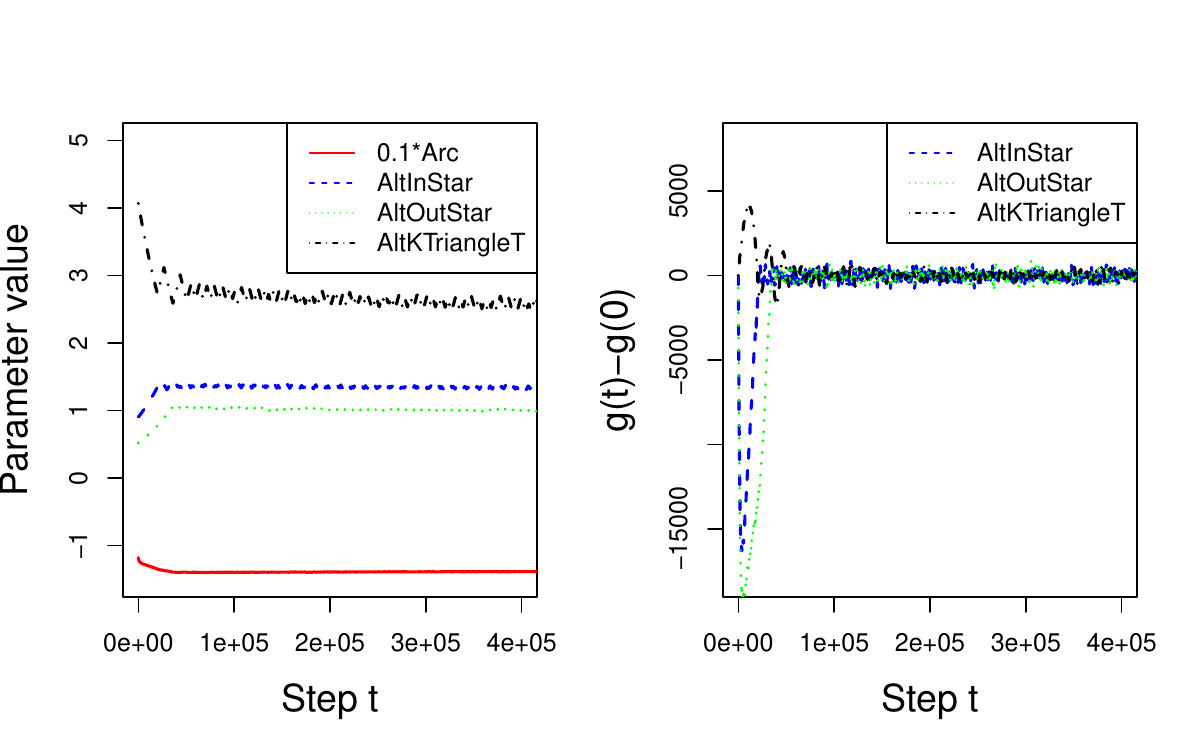}
	\caption{\label{Fig:5} ERGM parameters (a) and statistics $\bm{g}\left(x_{t}\right) - \bm {g}\left(x_{\rm{obs}}\right)$ (b) as a function of step $t$ of the EE algorithm. The starting parameter values are CD estimates.}
\end{figure*}

To demonstrate the scaling properties of the EE algorithm, given by~(\ref{eq:Max}), we applied it to estimate ERGM parameters of a large directed network. The ERGM was specified by Arc, popularity spread (AltInStar), activity spread (AltOutStar) and path closure (AltKTrianglesT) statistics, as detailed in~\cite{Lusher2013,Robins2009}. We fitted this ERGM specification to the who-trust-whom online social network Epinions, available from https://snap.stanford.edu/data/soc-Epinions1.html. The metaparameters of the EE algorithm were set to $a=0.00002$, $m=1000$, and we used the efficient IFD sampler~\cite{Byshkin2016}. Results of estimation of this empirical network with $75879$ nodes are given in \ref{Fig:5}. Producing these results took $7.4$ hours on a laptop.  The EE algorithm, proposed in this paper, is implemeted in the open-source software for the analysis of directed networks  \cite{Alex2019}, available from https://github.com/stivalaa/EstimNetDirected. Using this software we succesfully estimated ERGM parameters for many large-size social networks.

\section*{Conclusions and Outlook}

We propose an algorithm for Maximum Likelihood estimation of parameters of statistical models from the exponential family. The algorithm is based on combination of the method for maximum likelihood estimation \cite{Younes1988} and simulated annealing, a popular semiheuristic approach for multidimensional optimization \cite{Kirkpatrick1983}. Though these methods have different application areas, they have an interesting feature in common - they do not require equilibrium samples. The comparison of these methods and thier connection with statistical mechanics opens up new opportunities to design efficient methods for Monte Carlo inference.

Optimal parameter estimation is highly desirable, but not always possible. It is not always possible to obtain equilibrium samples of the distribution~(\ref{eq:1}), and the solution of the inverse problem, considered in this paper, requiers even larger advances. It is well known that Monte Carlo methods for statistical inference produce asymptotically exact results, but do not scale to big data~\cite{Blei2017}. Many approximate methods were developed to overcome this problem of scale, and the comparison of these methods would go far beyond of the scope of our reseach. Until recently, the largest network data for which MLE could  be computed was limited to few thousands of nodes~\cite{Krivitsky2017} and few millions of variables~\cite{Ghahramani2015}. The method we propose scales well to much larger data sets, and significantly increases these dimensions.

Another important advantage of the algorithm suggested in this paper is its simplicity. Because of its simplicity the algorithm can be easily incorporated into existing software that uses statistical models based on the Ising model, Markov Random Field, ERGMs or other models from the exponential family. These models are frequently adapted for the analysis of network data, and the algorithm described in this paper gives an effificent tool for the analysis of network data, that appears in many fields of science.

Currently we use the proposed approach to train restricted Boltzman machines \cite{Tieleman2008,Tramel2018} and adapt slightly different updating step (see Supplementary materials). The preliminary results are encouraging. In many important cases statistical models have latent variables and Younes suggested an extension of his algorithm to imperfect observations~\cite{Younes1991}.  Besides persistent contrastive devergence, Tieleman and Hinton proposed other popular algorithms for optimization in machine learning. These are prospective directions for future work.

\bibliography{scibib}

\bibliographystyle{Science}

\section*{Acknowledgments}
We thank Gerard Barkema, Christian P. Robert, Joris Bierkens, Ernst Wit and Antonietta Mira for helpful comments. We thank Alex Stivala for sharing the Estimnet-directed code. MB and AL thank Swiss National Science Foundation, grant number 167326, National Research Program 75 (“Big Data”) for financial support.


\newpage

\section*{Supplementary materials}


\subsection*{Starting point. Contrastive Divergence.} 

\paragraph{Theorem 1.} 
Let a transition probability $P(x\rightarrow x'| \bm\theta)$ define a Markov chain with a unique stationary distribution $\pi \left(x|\bm\theta \right)$.  Then for any $\bm{\theta}$, $\bm{g}(x)$ and $\pi \left(x|\bm\theta \right)$:
\begin{equation}
\sum_{x} \pi \left(x|\bm\theta \right)\Delta\bm{g}(x, \bm\theta) =0, \label{eq:T1}
\end{equation}
where
\begin{equation}
\Delta\bm{g}(x, \bm\theta)  =\sum_{x'} P(x\rightarrow x'| \bm\theta) \left[\bm g\left(x'\right) - \bm g\left(x\right)\right]. \label{eq:T2}
\end{equation}

\paragraph{Proof.}

\begin{equation}
\; \Delta\bm{g}(x, \bm\theta)  =\sum_{x'} P(x\rightarrow x'| \bm\theta)\bm g\left(x'\right) -  \bm g\left(x\right)\sum_{x'} P(x\rightarrow x'| \bm\theta),\nonumber
\end{equation}
\begin{equation}
\; \Delta\bm{g}(x, \bm\theta)  =\sum_{x'} P(x\rightarrow x'| \bm\theta)\bm g\left(x'\right) -  \bm g\left(x\right),\nonumber
\end{equation}
\begin{equation}
\; \sum_{x} \pi \left(x|\bm\theta \right)\Delta\bm{g}(x, \bm\theta)=\sum_{x,x'}\pi \left(x|\bm\theta \right) P(x\rightarrow x'| \bm\theta)\bm g\left(x'\right) - \sum_{x}  \pi \left(x|\bm\theta \right)\bm g\left(x\right). \nonumber
\end{equation}
\qquad \qquad  Using the global balance equation for Markov chains
\begin{equation}
\sum_{x}\pi \left(x|\bm\theta \right) P(x\rightarrow x'| \bm\theta)= \pi \left(x'|\bm\theta \right) \nonumber
\end{equation}
\qquad \qquad we obtain
\begin{equation}
\sum_{x} \pi \left(x|\bm\theta \right)\Delta\bm{g}(x, \bm\theta)= \sum_{x'}  \pi \left(x'|\bm\theta \right)\bm g\left(x'\right) - \sum_{x}  \pi \left(x|\bm\theta \right)\bm g\left(x\right)=0 \nonumber
\end{equation}

It is interesting to compare the result of this theorem with the results available in extant literature. The contrastive divergence (CD) estimator $\bm{\hat\theta}_{\rm{CD}}$ is a solution of the following equation \cite{Carreira-Perpinan2005,Byshkin2018} :
\begin{equation} \label{eq:cd}
\Delta \bm{g}\left(x_{\rm{obs}}, \bm{\hat\theta}_{\rm{CD}} \right)  =0.
\end{equation}
It is popular in machine learning and its value can be obtained by the algorithm proposed in \cite{Carreira-Perpinan2005}, described in the psedocode below. Consistency of the CD estimator is still under debate. However, Eq.~(\ref{eq:T1}) suggests that $\bm{\hat\theta}_{\rm{CD}}$ is a consistent estimator ~\cite{Byshkin2018}: if we have samples of $\pi \left(x|\bm{\theta^*} \right)$, the LHS of Eq.~(\ref{eq:T1}) can be computed by the Monte Carlo integration~\cite{Newman1999} and, using Eq.~(\ref{eq:T1}),  ${\bm{\theta^*}}$ can be found easily and fast.

Consider a Markov chain with transition probability $P(x\rightarrow x'| \bm\theta)$. Let at step $t=0$ we have $x_0=x_{\rm{obs}}$.  Then the expected statistics at step $t=1$  are given by
\begin{equation} \label{eq:non-eq0}
E_{\bm{\theta}}\left[\bm{g}(x_{1})\right]=\sum_{x'} P(x_{\rm{obs}}\rightarrow x'| \bm\theta)\bm g\left(x'\right) =\bm{g}(x_{\rm{obs}})  + \Delta\bm{g}(x_{\rm{obs}}, \bm\theta). 
\end{equation}
Hence $E_{\bm{\theta}}\left[\bm{g}(x_{1})\right]-\bm{g}(x_{\rm{obs}})=0$ if and only if $\Delta\bm{g}(x_{\rm{obs}},\bm\theta)=0$. And hence the solution of $\Delta\bm{g}(x_{\rm{obs}}, \bm\theta)=0$ is a good starting point for the solution of $E_{\bm{\theta}}\left[\bm{g}(x)\right]-\bm{g}(x_{\rm{obs}})=0$.

\subsection*{Pseudocode} 

\begin{table}[hbt!]
	\label{CD}
	\begin{tabular}{lll}
		\toprule
		\multicolumn{2}{l}{\textbf{Contrastive Divergence (CD) Algorithm}}                   \\
		\midrule
		1: & Initialization: $t = 0$ ; $x=x_{\rm{obs}}$; ${\bm\theta}_{0}=0$    \\
		2: & $\Delta\bm{g} = 0$ \\
		3: & \textbf{for} $k=1$ to $m$ (number of MC steps) \textbf{do} \\
		4: &  \hspace{5mm}Propose a trial move $x\rightarrow x'$ with probability $q\left(x\rightarrow x'\right)$ \\ 
		5: & \hspace{5mm} Calculate the Metropolis-Hastings acceptance probability $\alpha\left(x\rightarrow x', {\bm\theta}_{t}\right)$ \\
		6: & \hspace{5mm}\textbf{If} $Unif\left(\left [ 0, 1\right]\right) < \alpha\left(x\rightarrow x', {\bm\theta}_{t}\right)$ \textbf{then} $\Delta\bm{g} = \Delta\bm{g} + \bm{g}\left(x'\right) - \bm{g}\left(x\right)$ \\
		7: & \textbf{end for} \\
		8: & Update parameters ${\bm\theta}_{t+1}={\bm\theta}_{t}-a\cdot \Delta\bm{g} $ \\
		9: &  $t=t+1$; \textbf{If} $t<t_{\rm{CD}}$ \textbf{then} go to step 2  \textbf{else} $\bm{\hat{\theta}}_{\rm{CD}} = {\bm\theta}_{t}$  \\    
		\bottomrule
	\end{tabular}
\end{table}

\begin{table}[hbt!]
	\label{EE}
	\begin{tabular}{lll}
		\toprule
		\multicolumn{2}{l}{\textbf{Equilibrium Expectation (EE) Algorithm}}                   \\
		\midrule
		1: & Initialization: $t = 0$ ; $x=x_{\rm{obs}}$; ${\bm\theta}_{0}= \bm{\hat{\theta}}_{\rm{CD}}$ ; $\Delta\bm{g} = 0$   \\
		
		2: & \textbf{for} $k=1$ to $m$ (number of MC steps) \textbf{do} \\
		3: & \hspace{5mm}Propose a trial move $x\rightarrow x'$ with probability $q\left(x\rightarrow x'\right)$ \\ 
		4: & \hspace{5mm}Calculate the Metropolis-Hastings acceptance probability $\alpha\left(x\rightarrow x', {\bm\theta}_{t}\right)$ \\
		5: & \hspace{5mm}\textbf{If} $Unif\left(\left [ 0, 1\right]\right) < \alpha\left(x\rightarrow x', {\bm\theta}_{t}\right)$ \textbf{then} $\Delta\bm{g} = \Delta\bm{g} + \bm{g}\left(x'\right) - \bm{g}\left(x\right)$ \\
		& \hspace{15mm} \textbf{and perform this move:} $x=x'$\\
		6: & \textbf{end for} \\
		7: & Update parameters  ${\bm\theta}_{t+1}={\bm\theta}_{t}-a\cdot \max \left(\left| {\bm\theta}_{t} \right|, c\right)\cdot \mathrm{sign}(\Delta\bm{g} )$ \\
		8: & $t=t+1$; Save sequences $\Delta\bm{g}_{t}=\Delta\bm{g}$; \\
		&   \textbf{If} $t<t_{\rm{EE}}$ \textbf{then} go to step 2  \textbf{else}  
		$ \bm{\hat\theta}_{\rm{MLE}}=\frac{1}{t-t_B}\sum_{j=t_B+1}^{t}{\bm\theta}_j$ \\    
		\bottomrule
	\end{tabular}
\end{table}

\newpage
\subsection*{Relation with previous work}

In the algorithm, poposed in ~\cite{Byshkin2018}, the step size was a function of $\bm g\left(x_{\rm{obs}}\right)-\bm g\left(x_{t}\right)$  too, but the learning rate was adapted so that
\begin{equation} \label{eq:cond}
\sigma(\theta_{i})\approx A\cdot \max\left(\left|\left \langle {\theta}_{i} \right \rangle \right|, c\right) \forall i, 
\end{equation}
where $\left \langle .. \right \rangle$ is the averaging over the current states of the Markov chain,  $\sigma(..)$ is a standard deviation over these states, and $A$  is a positive constant that moderates the learning rate. We believe that by $\sigma({\theta}_{i})$  the step size was measured and the step size was adapted so that it was proportional to $\left \langle {\theta}_{i} \right \rangle$. Intensive tests were performed on many different ERGM specifications and datasets. It was reported that the algorithm converges faster when the approximate condition~(\ref{eq:cond}) is satisfied, and that when~(\ref{eq:cond}) is not satisfied the algorithm may often not converge. A complicated adaptive method was applied to satisfy~(\ref{eq:cond}). 

Computational experiments show that when the EE algoritm proposed in this paper is applied, the approximate equality~(\ref{eq:cond}) is satisfied. 
Thus the algorithm proposed in this paper may be considered as a simple version of the adaptive EE algorithm~\cite{Byshkin2018}. 

\subsection*{Convergence criterion} 
\paragraph{}
To understand why and when the EE algorithm converges to MLE, it is useful  to consider the following convergence test: I)  ${\bm\theta}_{t}$ converges  and  II) the  t-ratio test given by Eq. (12) in the main text:
\begin{equation} \label{eq:t-ratio}
\frac{\left| \left \langle g_{i}\left(x_{t}\right)- g_{i}\left(x_{\rm{obs}}\right)\right \rangle \right|}{\sigma\left[g_{i}\left(x_{t}\right)- g_{i}\left(x_{\rm{obs}}\right)\right]} <\tau~  \forall i,
\end{equation}
where  $\left \langle .. \right \rangle$ and $\sigma(..)$ is the mean and the standard deviation over  $t>t_B$ ,  $\tau=0.1$. 
If ${\bm\theta}_{t}$  converges to a constant ${\bm{\bar\theta}}$ then $x_{t}$ follows $\pi \left(x|{\bm{\bar\theta}} \right)$ and $\left \langle \bm {g}\left(x_{t}\right) \right \rangle$  is a Monte Carlo estimator of $E_{{\bm{\bar\theta}}}\bm g\left(x\right)$. In this case  Eq.~(\ref{eq:t-ratio}) is a robust criterion that ${\bm{\bar\theta}}$ is MLE ~\cite{Snijders2002}.  From Fig. 1 in the main text  we see that, after ${\bm\theta}_{t}$  converges, it fluctuates around ${\bm{\bar\theta}}$ , that we compute by averaging. Are we sure that ${\bm{\bar\theta}}$ is MLE? When the EE algorithm is applied, the parameters are not constant, but fluctuate, and hence we sample from $\pi \left(x|\bm{\theta} \right)$  with uncertain  $\bm {\theta}$. 

Ceperley and Dewing~\cite{Ceperley1999} proposed the MCMC algorithm to sample from  $\pi \left(x|\bm{\theta} \right)$ with uncertain energy. Recently, the result of Ceperley and Dewing was rederived by Frenkel and coauthors~\cite{Frenkel2017} utilizing a more general method of stochastic weight functions  and we will use this result here: if fluctuation of $\beta \left[ E\left(x', \bm{\theta}\right) - E\left(x, \bm{\theta}\right) \right]$  is normally distributed with variance $\sigma$  (the central limit theorem guarantees it if averaging is long enough), then, assuming symmetric proposals for simplicity, the acceptance probability, that samples from the correct $\pi \left(x|\bm{\bar\theta} \right)$, becomes
\begin{equation} \label{eq:Frenkel}
\alpha(x\rightarrow x', \bm{\theta})=\min \left\{1,\exp \left[-\beta\left( E\left(x', \bm{\theta}\right) - E\left(x, \bm{\theta}\right) \right) - \frac{\sigma^{2}}{2}\right]\right\}.
\end{equation}
In the EE algorithm, the parameters $\bm{\theta}$ are not constant, but fluctuate, imposing noise on \newline
$\beta \left[ E\left(x', \bm{\theta}\right)-E\left(x,\bm{\theta}\right) \right])$. Hence, the acceptance probability with the penalty term of Ceperley and Dewing must be used with the EE algorithm.

In practice, the penalty term $\sigma^{2}/2$ may be small and can be neglected. For simplicity, consider the Ising model with one parameter again. The standard deviation $\sigma_{\theta}$  of this parameter is given by Eq.~(\ref{eq:cond}) and hence, assuming $\left|\bar\theta \right|>c$, one obtains that $\sigma^2 \propto A^2$. Computational experiments show that  $A\propto a$, hence $\sigma^2\propto a^2 $ and we see that the learning rate $a$ can be selected small enough so that the penalty term $\sigma^{2}/2$ is small and can be neglected. And if the penalty term $\sigma^{2}/2$  can be neglected, then $\bm{\bar \theta}$  is MLE and the uncertainty of MLE, computed in this way, is given by Eq.~(\ref{eq:cond}). Thus we have shown that, if the EE algorithm converges, it converges to MLE.

In the EE algorithm it is crucial that the step size is small with respect to the corresponging parameter value.  For instance, the following updating step may be appropriate for both Exponential Random Graph models and Restricted Boltzman Machines:
\begin{equation} \label{eq:EEhidden}
{\bm\theta}_{t+1}={\bm\theta}_{t}+a\cdot  \left(\left| {\bm\theta}_{t} \right|+ c\right)\cdot\left[\bm g\left(x_{0}\right) - \bm g\left(x_{t+1}\right)\right].
\end{equation}


\end{document}